\documentclass[manuscript]{aastex}
\DeclareTextSymbol{\degre}{T1}{6}
\DeclareTextSymbol{\degre}{OT1}{23}
\usepackage{wasysym}
\usepackage{color}

\slugcomment{Submitted to The Astrophysical Journal Letters}

\shorttitle{The dual origin of the nitrogen deficiency in comets}
\shortauthors{O. Mousis et al.}

\begin{document}


\title{The dual origin of the nitrogen deficiency in comets: selective volatile trapping in the nebula and postaccretion radiogenic heating\\
}


 \author{
Olivier~Mousis\altaffilmark{1,*},
Aur{\'e}lie Guilbert-Lepoutre,\altaffilmark{2},
Jonathan~I.~Lunine\altaffilmark{3},
Anita~L.~Cochran\altaffilmark{4},
J.~Hunter~Waite\altaffilmark{5},
Jean-Marc Petit\altaffilmark{1},
Philippe Rousselot\altaffilmark{1}}

\altaffiltext{1}{Universit{\'e} de Franche-Comt{\'e}, Institut UTINAM, CNRS/INSU, UMR 6213, Observatoire des Sciences de l'Univers de Besancon, France}

\altaffiltext{*}{Corresponding author E-mail address: olivier.mousis@obs-besancon.fr}

\altaffiltext{2}{Department of Earth and Space Sciences, UCLA, Los Angeles, CA 90095, USA}
 
\altaffiltext{3}{Center for Radiophysics and Space Research, Space Sciences Building Cornell University,  Ithaca, NY 14853, USA}

\altaffiltext{4}{The University of Texas McDonald Observatory, Austin, TX 78712 USA}

\altaffiltext{5}{Space Science and Engineering Division, Southwest Research Institute, San Antonio, Texas, USA}

\begin{abstract}

We propose a scenario that explains the apparent nitrogen deficiency in comets in a way consistent {\bf with the fact that the surfaces of Pluto and Triton are dominated by nitrogen-rich ice}. We use a statistical thermodynamic model to investigate the composition of the successive multiple guest clathrates that may have formed during the cooling of the primordial nebula from the most abundant volatiles present in the gas phase. These clathrates agglomerated with the other ices (pure condensates or stoichiometric hydrates) and formed the building blocks of comets. We report that molecular nitrogen is a poor clathrate former, when we consider a plausible gas phase composition of the primordial nebula. This implies that its trapping into cometesimals requires a low disk temperature ($\sim$20 K) in order to allow the formation of its pure condensate. We find that it is possible to explain the lack of molecular nitrogen in comets as a consequence of their postformation internal heating engendered by the decay of {\bf short-lived radiogenic nuclides}. This scenario is found consistent with the presence of {\bf nitrogen-rich ice covers on Pluto and Triton}. Our model predicts that comets should present xenon-to-water and krypton-to-water ratios close to solar xenon-to-oxygen and krypton-to-oxygen ratios, respectively. In contrast,  the argon-to-water ratio is predicted to be depleted by a factor of $\sim$300 in comets compared to solar argon-to-oxygen, as a consequence {\bf of poor trapping efficiency and radiogenic heating}.

\end{abstract}





\section{Introduction}

The abundance of the super volatile molecule N$_2$ in cometary nuclei is still under debate. Searches for UV fluorescence from N$_2$ with the Far Ultraviolet Spectroscopic Explorer (FUSE) have been unsuccessful (Bockel\'ee-Morvan et al. 2004). Measurements of the N$_2$ abundance in 1P/Halley by Giotto were not feasible: owing to the low resolution of the mass spectrometer, it was not possible to discriminate N$_2$ from CO (Eberhardt et al. 1987). Ground-based observation of the N$_2^+$ band at 3914~\AA~is a difficult challenge due both to the contamination of the $C_3$ (0,2,0)-(0,0,0) band and the presence of telluric N$_2^+$ emission lines. Different positive detections of this feature have been claimed in the ionic tail of comets Humason (1962 VIII) (Greenstein \& Arpigny 1962), 1P/Halley (Lutz et al. 1993; Wyckoff \& Theobald 1989), C/1987 P1 Bradfield (Lutz et al. 1993), and C/2002 VQ94 (LINEAR) (Korsun et al. 2006). Arpigny, from photographic spectra and other archival data estimated a positive N$_2^+$/CO$^+$ intensity ratio for 12 comets (Cochran et al. 2000). All these positive detections are based on low-resolution spectra. So far similar searches with high-resolution spectra have been unsuccessful. This was the case for observations of comets 122P/de Vico, C/1995 O1 (Hale-Bopp), and 153P/2002 C1 (Ikeya-Zhang) that did not detect N$_2^+$ band, yielding upper limits of $\le$ 10$^{-5}$--10$^{-4}$ on the abundance of N$_2$ relative to CO (Cochran et al. 2000; Cochran 2002).

An interpretation of the possible N$_2$ deficiency in comets has been proposed by Iro et al. (2003). These authors assumed that comets were made of clathrate hydrates (hereafter clathrates) and argued that CO was preferentially incorporated in clathrates compared to N$_2$ after having investigated the competition between the trapping of these two molecules in clathrates. In order to explain the N$_2$ deficiency in comets, they argued that the nebula's temperature never cooled down below $\sim$45 K in their formation region, impeding the formation of N$_2$-bearing ice (in clathrate or pure condensate form) that would have been agglomerated by comets. However, this mechanism is not consistent with the fact that Pluto and Triton {\bf possess thick nitrogen ice covers} (Lellouch et al. 2011) while they are both expected to have been formed in the same region of the primitive nebula as ecliptic comets (Kavelaars et al. 2001). In addition, Iro et al. (2003) did not consider the competition between the trapping of N$_2$ in clathrates with that of the other abundant volatiles present in the nebula (CO, CO$_2$, CH$_4$, H$_2$S, ...). Indeed, due to their various propensities for trapping, the consideration of a larger set of molecules can drastically change the calculated fraction of N$_2$ incorporated in clathrates formed in the nebula.

Here we propose a scenario that could explain the apparent N$_2$ deficiency in comets in a way consistent with the presence of this molecule {\bf on the surfaces of Pluto and Triton}. We use a statistical thermodynamic model to investigate the composition of the successive multiple guest clathrates (hereafter MG clathrates) that may have formed during the cooling of the primordial nebula from the most abundant volatiles present in the gas phase (see section 2). These clathrates agglomerated with the other ices and formed the building blocks of comets (hereafter cometesimals). The major ingredient of our model is the description of the guest-clathrate interaction by a spherically averaged Kihara potential with a nominal set of potential parameters. Our model allows us to report that N$_2$ is a poor clathrate former when considering a plausible gas phase composition of the primordial nebula, implying that its trapping into cometesimals requires a low disk's temperature ($\sim$20 K) in order to allow the formation of its pure condensate. We find that it is possible to explain the lack of nitrogen in comets as a consequence of their postaccretion internal heating engendered by the decay of radiogenic nuclides (see section 3). 

\section{Composition of cometesimals formed in the primordial nebula}
\label{composition}

\subsection{Modeling approach}

{In our model, the volatile phase incorporated in cometesimals is composed of a {\bf mixture} of pure ices, stoichiometric hydrates (such as {\bf NH$_3$-H$_2$O hydrate}) and MG clathrates that crystallized in the form of microscopic grains at various temperatures in the outer part of the disk.  Our model is based on the assumption that cometesimals have grown from the agglomeration of these icy grains due to collisional coagulation, implying no loss of volatile during their growth phase (Weidenschilling 1997).} Here, the clathration process stops when no more crystalline water ice is available to trap the volatile species and then only pure condensates form at lower temperature in the disk. The process of volatiles trapping in {icy grains} is calculated using the equilibrium curves of hydrates and pure condensates, our model determining the equilibrium curves and compositions of MG clathrates, and the thermodynamic paths detailing the evolution of temperature and pressure between 5 and 30 AU in the protoplanetary disk. We refer to the work of Alibert et al. (2005) for a full description of the turbulent model of accretion disk used here.

The composition of the initial gas phase of the disk is defined as follows: we assume that the abundances of all elements (C, N, O, S, P, Ar, Kr and Xe) are protosolar (Asplund et al. 2009) and that O, C, and N exist only under the form of H$_2$O, CO, CO$_2$, CH$_3$OH, CH$_4$, N$_2$, and NH$_3$. The abundances of CO, CO$_2$, CH$_3$OH, CH$_4$, N$_2$ and NH$_3$ are then determined from the adopted CO:CO$_2$:CH$_3$OH:CH$_4$ and N$_2$:NH$_3$ gas phase molecular ratios. Once the abundances of these molecules are fixed, the remaining O gives the abundance of H$_2$O. Concerning the distribution of elements in the main volatile molecules, we set CO:CO$_2$:CH$_3$OH:CH$_4$~=~70:10:2:1 in the gas phase of the disk, values that are consistent with the ISM measurements considering the contributions of both gas and solid phases in the lines of sight (Mousis et al. 2009). In addition, S is assumed to exist in the form of H$_2$S, with H$_2$S:H$_2$~=~0.5 $\times$(S/H$_2$)$_{\odot}$, and other refractory sulfide components. We also consider N$_2$:NH$_3$~=~1:1 in the nebula gas-phase, a value compatible with thermochemical calculations in the solar nebula that take into account catalytic effects of Fe grains on the kinetics of N$_2$ to NH$_3$ conversion (Fegley 2000). Note that CH$_3$OH is considered to be formed only as a pure condensate in our calculations since, to the best of our knowledge, no experimental data concerning the equilibrium curve of its associated clathrate have been reported in the literature.

The equilibrium pressure for the successive MG clathrates formed in the solar nebula can be expressed as (Hand et al. 2006):

\begin{equation}
P_{eq,MG} = \left [\sum_i \frac{y_i}{P_{eq,i}} \right ]^{-1}
\end{equation}

\noindent where $y_i$ is the mole fraction of the component $i$ in the fluid phase. The equilibrium pressure curves of each species are determined by fitting the available theoretical and laboratory data (Lunine \& Stevenson 1985). Their equations are of the form $ \log P_{eq,i} = A/T + B$, where $P_{eq,i}$ and $T$ are the partial equilibrium pressure (bars) and temperature (K) of the considered species $i$, respectively.

The relative abundances of guest species incorporated in MG clathrate formed at a given temperature and pressure from the solar nebula gas phase are calculated following the method described by Lunine \& Stevenson (1985) and Mousis et al. (2010), which uses classical statistical mechanics to relate the macroscopic thermodynamic properties of clathrates to the molecular structure and interaction energies. It is based on the original ideas of van der Waals \& Platteeuw (1959) for clathrate formation, which assume that trapping of guest molecules into cages corresponds to the three-dimensional generalization of ideal localized adsorption. This approach is based on four key assumptions (Lunine \& Stevenson 1985; Sloan 1998):

\begin{enumerate}
\item The host molecules contribution to the free energy is independent of the clathrate occupancy. This assumption implies in particular that the guest species do not distort the cages.

\item (a) The cages are singly occupied. (b) Guest molecules rotate freely within the cage.

\item Guest molecules do not interact with each other.

\item Classical statistics is valid, i.e., quantum effects are negligible.

\end{enumerate}

In this formalism, the fractional occupancy of a guest molecule $K$ for a given type $t$ ($t$~=~small or large) of cage can be written as

\begin{equation}
\label{occupation}
y_{K,t}=\frac{C_{K,t}P_K}{1+\sum_{J}C_{J,t}P_J} ,
\end{equation}

\noindent where the sum in the denominator includes all the species which are present in the initial gas phase. $C_{K,t}$ is the Langmuir constant of species $K$ in the cage of type $t$, and $P_K$ is the partial pressure of species $K$. This partial pressure is given by $P_K=x_K\times P$ (we assume that the sample behaves as an ideal gas), with $x_K$ the mole fraction of species $K$ in the gas phase, and $P$ the total gas pressure, which is dominated by H$_2$. 

The Langmuir constant depends on the strength of the interaction between each guest species and each type of cage, and can be determined by integrating the molecular potential within the cavity as

\begin{equation}
\label{langmuir}
C_{K,t}=\frac{4\pi}{k_B T}\int_{0}^{R_c}\exp\Big(-\frac{w_{K,t}(r)}{k_B T}\Big)r^2dr ,
\end{equation}

\noindent where $R_c$ represents the radius of the cavity assumed to be spherical, $k_B$ the Boltzmann constant, and $w_{K,t}(r)$ is the spherically averaged Kihara potential representing the interactions between the guest molecules $K$ and the H$_2$O molecules forming the surrounding cage $t$. For a spherical guest molecule, this potential $w(r)$ can be written as (McKoy \& Sinano\u{g}lu 1963):

\begin{equation}
\label{pot_Kihara}
w(r) = 2z\epsilon\Big[\frac{\sigma^{12}}{R_c^{11}r}\Big(\delta^{10}(r)+\frac{a}{R_c}\delta^{11}(r)\Big) - \frac{\sigma^6}{R_c^5r}\Big(\delta^4(r)+\frac{a}{R_c}\delta^5(r)\Big)\Big],
\end{equation}

\noindent with

\begin{equation}
\delta^N(r)=\frac{1}{N}\Big[\Big(1-\frac{r}{R_c}-\frac{a}{R_c}\Big)^{-N}-\Big(1+\frac{r}{R_c}-\frac{a}{R_c}\Big)^{-N}\Big].
\end{equation}

\noindent In Eq. (\ref{pot_Kihara}), $z$ is the coordination number of the cell. This parameter depends on the structure of the clathrate (I or II) and on the type of the cage (small or large). The Kihara parameters $a$, $\sigma$ and $\epsilon$ for the molecule-water interactions employed in this work are listed in Table \ref{kihara}. Note that our parameters are different from those used by Iro et al. (2003), who instead of using potential parameters describing the guest-clathrate interaction in their statistical model, employed potential parameters corresponding to guest--guest interactions (case of pure solutions). {When comparing different data sets determined for the same X-H$_2$O interaction (with X = CH$_4$, N$_2$ and Xe), we opted for the most recent ones because they are fitted to a larger range of experimental data. In the case of Ar, Kr, CO and PH$_3$, the listed sets are the unique ones that we found in the literature. For the H$_2$S-H$_2$O and CO$_2$-H$_2$O interaction parameters we used those of Parrish and Prausnitz (1972) and Thomas et al. (2009) (the set of CO$_2$-H$_2$O interaction parameters provided by Thomas et al. (2009) is an update of the one determined by Parrish and Prausnitz (1972) and very close to the one recently derived by Herri and Chassefi{\`e}re (2012)). We also investigated the interaction parameters of Sloan and Koh (2008) but these parameters yield unphysical results when used in our models.  We suspect this is because the code of Sloan and Koh (2008) is optimized for industrial purposes and the parameters are not appropriate for our temperature-pressure regime.  We have tested our code against the one of Herri and Chassefi{\`e}re (2012) so we do not believe the problem resides in our code.} Finally, the mole fraction $f_K$ of a guest molecule $K$ in a clathrate can be calculated with respect to the whole set of species considered in the system as

\begin{equation}
\label{abondance} f_K=\frac{b_s y_{K,s}+b_\ell y_{K,\ell}}{b_s \sum_J{y_{J,s}}+b_\ell \sum_J{y_{J,\ell}}},
\end{equation}

\noindent where $b_s$ and $b_l$ are the number of small and large cages per unit cell respectively, for the clathrate structure under consideration, and with $\sum_K$ $f_K~=~1$. Values of $z$, $R_c$, $b_s$ and $b_l$ are taken from Parrish \& Prausnitz (1972). Each time a species is fully trapped in a MG clathrate, its gaseous abundance is set to zero in the disk. This change of the disk's gas phase composition induces the formation of a new MG clathrate at different temperature and pressure conditions and with a distinct composition. As a consequence, the compositions and formation conditions of successive MG clathrates are calculated $n$ times in the nebula, with $n$ corresponding to the total number of fully enclathrated species. Once all the water budget has been used for clathration, the volatiles remaining in the gas phase form pure condensates at lower temperatures. The equilibrium curves of these pure condensates derive from the compilation of laboratory data given in the CRC Handbook of Chemistry and Physics (Lide 2002).

The composition of the volatile phase incorporated into cometesimals formed at a given distance from the Sun is finally given by the intersection of the disk's thermodynamic path at this location with the equilibrium curves of the different formed ices. The amount of volatile, $i$ (relative to water), used to form the ice $j$ in the nebula (either in the form of stoichiometric hydrate, MG clathrate or of pure condensate), can be determined by the relation:

\begin{equation}
{m_{i,j} =  \frac{X_{i,j}}{X_{H_2O}} \frac{\Sigma(r; T_j, P_j)}{\Sigma(r; T_{H_2O}, P_{H_2O})}},
\end{equation}

\noindent where $X_{i,j}$ is the mass mixing ratio with respect to H$_2$ of the volatile $i$ used to form ice $j$ in the nebula. $X_{H_2O}$ is the mass mixing ratio of H$_2$O with respect to H$_2$ in the nebula. $\Sigma(R; T_j, P_j)$ and $\Sigma(R; T_{H_2O}, P_{H_2O})$ are the surface density of the disk at distance $R$ from the Sun at the epoch of formation of ice $j$, and at the epoch of condensation of water, respectively. The global volatile, $i$, to water mass ratio incorporated in cometesimals is then given by $M_j = \sum_{j=1,k} m_j$, with $k$ corresponding to the number of solid phases in which volatile $i$ is incorporated. Note that a thermodynamic path has been arbitrarily selected at the heliocentric distance of 5 AU to determine the composition of the formed ices. The adoption of any other distance range between $\sim$5 and 30 AU in the nebula would not affect the composition of the ices because it remains almost identical irrespective of i) their formation distance and ii) the input parameters of the disk, provided that the initial gas phase composition is homogeneous (Marboeuf et al. 2008).

\subsection{Results}

\begin{figure}
\begin{center}
\resizebox{\hsize}{!}{\includegraphics{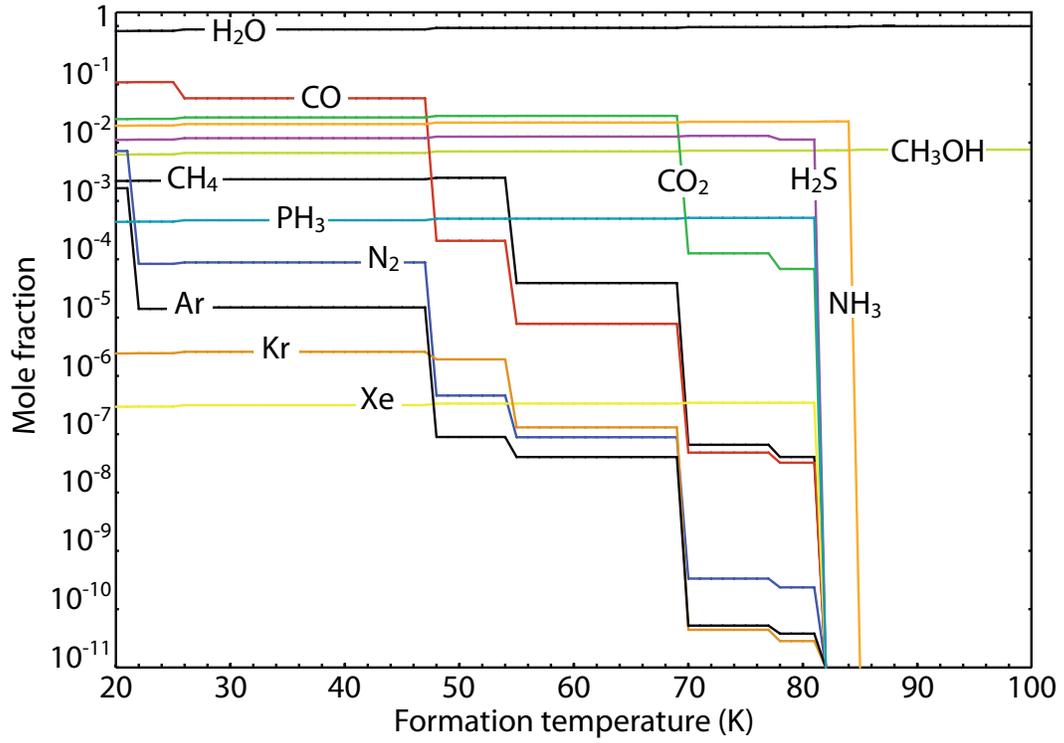}}
\caption{Composition of the volatile phase incorporated in cometesimals as a function of their formation temperature in the outer solar nebula.} 
\label{mole}
\end{center}
\end{figure}

\begin{figure}
\begin{center}
\resizebox{\hsize}{!}{\includegraphics{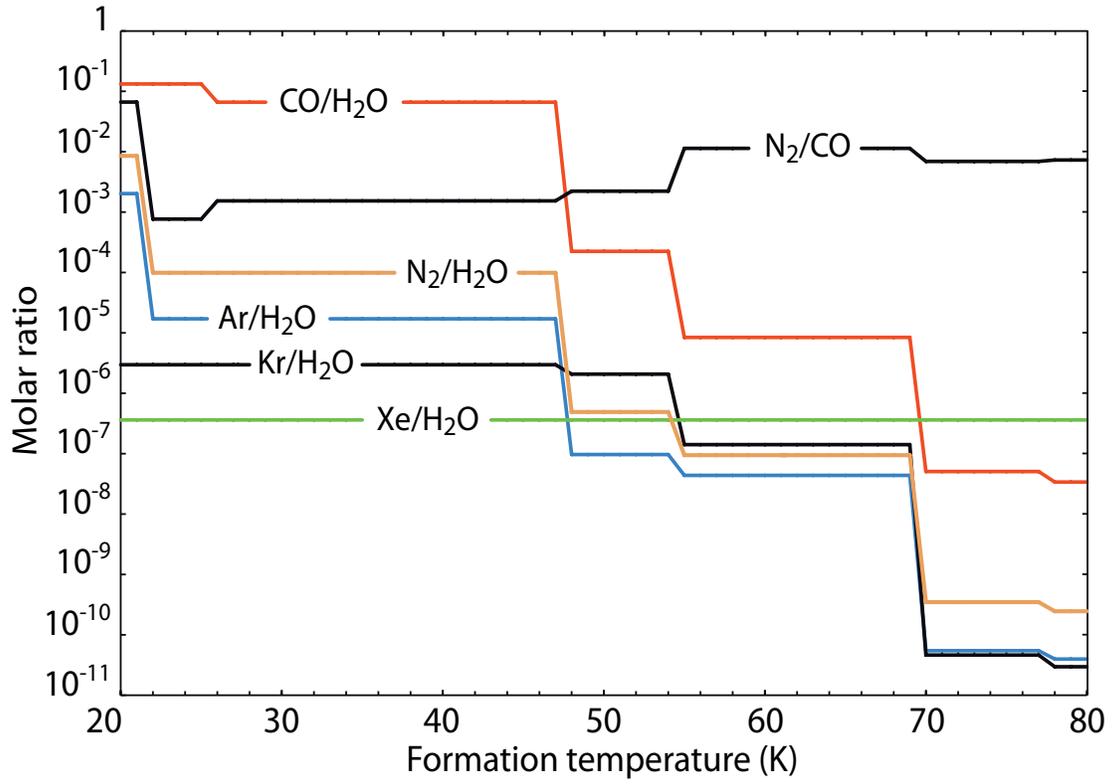}}
\caption{Molar ratios between different volatiles calculated in cometesimals as a function of their formation temperature.} 
\label{ratios}
\end{center}
\end{figure}

Figure \ref{mole} represents the composition of the volatile phase incorporated in {cometesimals agglomerated from ices} condensed in the outer nebula and Figure \ref{ratios} displays the molar ratios between incorporated volatiles of interest. Both figures show calculations expressed as a function of the formation temperature of cometesimals in the primordial nebula. {Observations of active comets when their activity is driven by the sublimation of water ice -- i.e. for heliocentric distances smaller than about 2~AU -- suggest that the mean CO production rate relative to water is typically a few percents, with extreme values comprised between 0.4 and 30\% (Bockel{\'e}e-Morvan et al. 2004).} Assuming that the production rates relative to water measured in comets are representative of the bulk composition of cometesimals, Figure \ref{ratios} shows that their formation temperature must be lower than $\sim$47 K in the nebula to get CO/H$_2$O $>$ 1\% in their interior. The figure also shows that a formation temperature lower than $\sim$25K would increase the CO/H$_2$O ratio up to $\sim$$13\%$ in cometesimals, as a result of the incorporation of pure CO condensate at this lower nebular temperature. 

{Formation temperatures of ices lower than $\sim$22 K in the nebula allow trapping of N$_2$ with N$_2$/H$_2$O of order of 1\% in the interior of planetesimals}, a value far above the highest estimates ($\sim$0.02\%) inferred from observations (Wyckoff et al. 1991; Cochran et al. 2000). On the other hand, assuming a formation temperature of cometesimals in the 22--47 K range in the nebula would allow N$_2$/H$_2$O $\sim$0.01$\%$ and  CO/H$_2$O $\sim$6.6$\%$ in their interior, values that are consistent with observations. A simple explanation for the formation of cometesimals would then argue that {they need to be agglomerated from grains formed in the 22--47 K range} in order to match the observations of the cometary production rates. However, this mechanism is not consistent with the fact that Pluto and Triton possess {\bf N$_2$-rich ice covers} while they are expected to be formed in the same region as comets. An alternative possibility would be that comets accreted from planetesimals formed at very low temperature in the disk, but that subsequent radiogenic heating enabled their partial devolatilization through their pores. In the next Section, we explore how radiogenic heating can induce important losses of N$_2$ in comets after their formation.

\section{Thermal evolution of comets}
\label{thermal}
\subsection{Nucleus model}
\label{nucleus}

\begin{figure}
\begin{center}
\resizebox{\hsize}{!}{\includegraphics{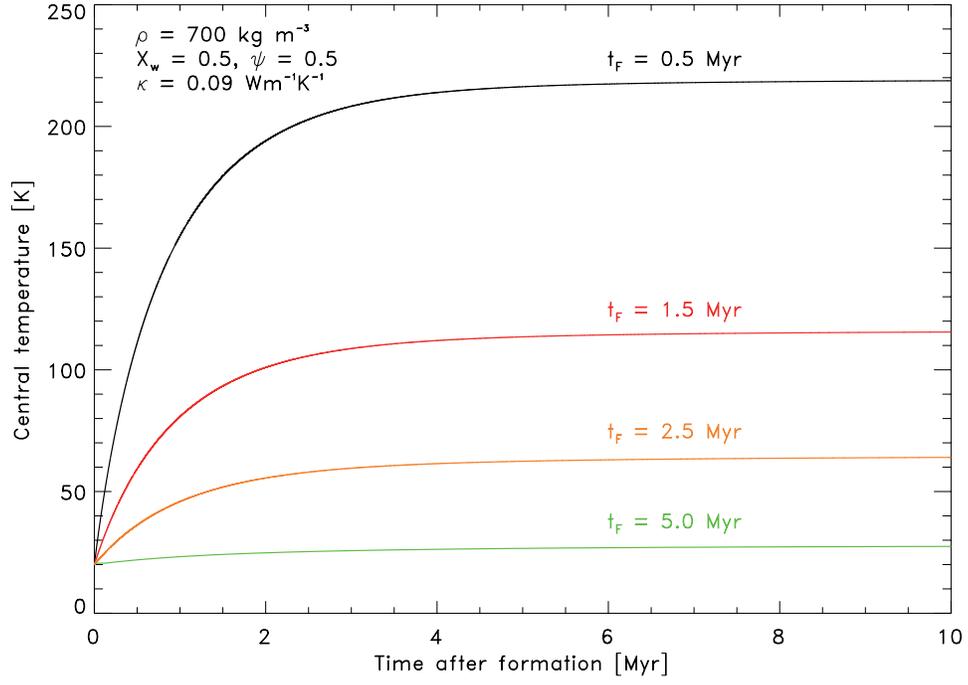}}
\caption{Time evolution of central temperature calculated fror different formation times.} 
\label{formation}
\end{center}
\end{figure}

\begin{figure}
\begin{center}
\includegraphics[scale=0.7]{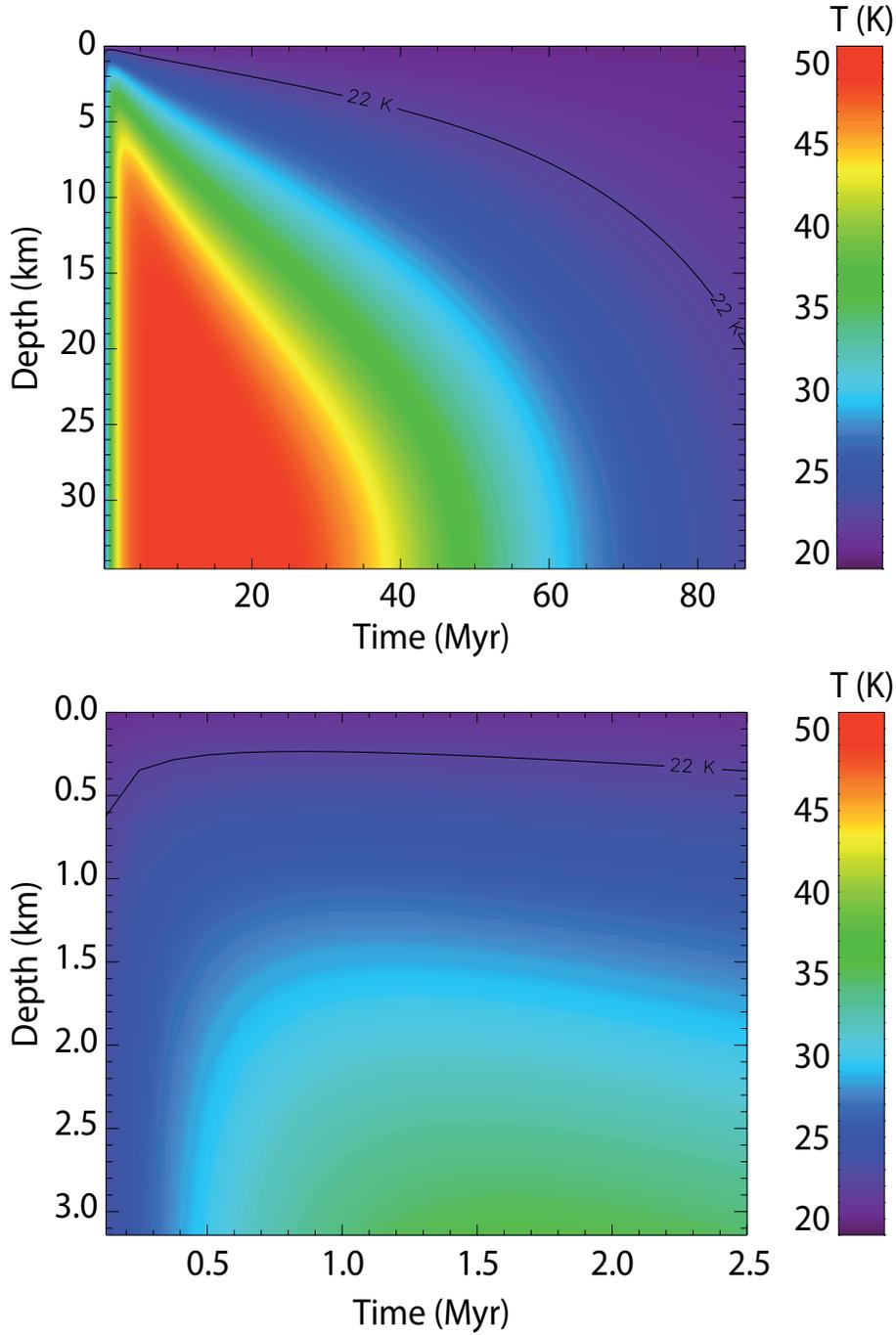}
\caption{Time evolution of the temperature profile within the body after a formation delay of 3~Myr. The parameters of the model are those adopted in Fig. \ref{formation}. Top panel represents a global view of the body's heating due to radionuclides and then cooling over a timespan of 85 Myr after formation. Bottom panel represents a zoom of the temperature evolution of the body's subsurface over 2.5 Myr after formation.} 
\label{profile}
\end{center}
\end{figure}

\begin{figure}
\begin{center}
\resizebox{\hsize}{!}{\includegraphics{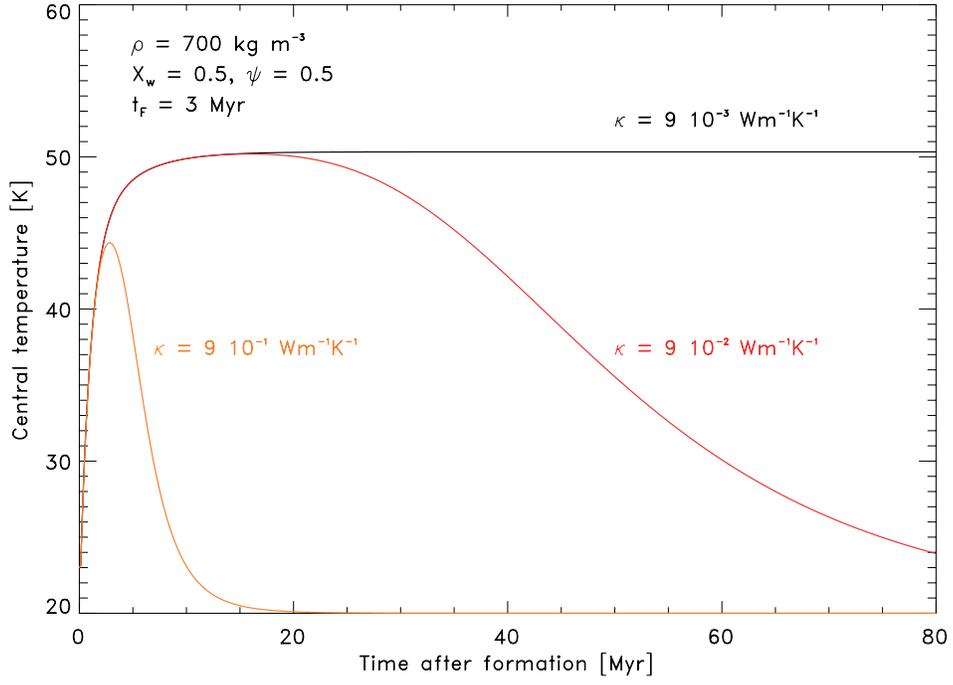}}
\caption{Time evolution of central temperature after a formation delay of 3~Myr, calculated for different values of the thermal conductivity.} 
\label{conductivity}
\end{center}
\end{figure}

{We use a fully three dimensional model of heat transport described by \citet{Gui11} to compute the time evolution of the temperature distribution within cometary nuclei, accounting for the heating due to the radioactive decay of short-lived nuclides ${}^{26}$Al and ${}^{60}$Fe. We assume that these objects formed by cold accretion, i.e. no accretional heating is accounted for. The model computes the temperature distribution and its evolution with time, by solving the heat diffusion equation:

\begin{equation}
\label{eq_depart}
\rho _{bulk}c~\frac{\partial T}{\partial t}~+~\nabla(- \kappa \overrightarrow{\nabla}T)~ =\mathcal{Q}_{rad}
\end{equation}

\noindent with $\rho _{bulk}$ [kg m$^{-3}$] the bulk density, $c$ [J kg$^{-1}$ K$^{-1}$] the heat capacity, $\kappa$ [W m$^{-1}$ K$^{-1}$] the thermal conductivity and $\mathcal{Q}_{rad}$ [W m$^{-3}$] the internal power production per unit volume due to the decay of short-lived radioactive nuclides. As an example, we show in this paper the results for an idealized ``Hale-Bopp" with a size of 35~km \citep{Sza11, Lam04, Fer03, Wea97}. We consider this object as a sphere, made of a porous mixture of crystalline water ice and dust, homogeneously distributed within the ice matrix. We use generic parameters to describe the bulk material: we consider it can be described with a dust to water ice mass ratio $X_{d}/X_w$ = 1 (with $X_d$ and $X_w$ the mass fractions of dust and water respectively), a porosity $\psi$ = 50\%, and a bulk density $\rho _{bulk}$ = 700~kg~m$^{-3}$ (typical values of the comet literature; see \cite{Hue06}). The heat capacity and thermal conductivity of such a mixture are $c$ = 800~J~kg$^{-1}$~K$^{-1}$ and $\kappa$ = 0.09~W~m$^{-1}$~K$^{-1}$ respectively. The thermal evolution is computed with the formation time $t_F$ --time for an object to grow to its final size-- as a reference for time zero, from an initial temperature of 20~K.

Heating due to radioactive decay is described by:
\begin{equation}
\mathcal{Q}_{rad}=~\sum _{rad}\rho _{d} ~ X_{rad}(t_F)~ H_{rad}\frac{1}{ \tau _{rad}} \exp \left( \frac{-t}{\tau _{rad}}\right),
\end{equation}
with $\rho _{d}$ the dust bulk density, $X_{rad}(t_F)$ the mass fraction of the given radioactive isotope after a formation time $t_F$, $H_{rad}$ the heat released per unit mass upon decay (4.84~$\times$~10$^{12}$~J~kg$^{-1}$ and 5.04~$\times$~10$^{12}$~J~kg$^{-1}$ for ${}^{26}$Al and ${}^{60}$Fe respectively \citep[see][]{Gui11}, $\tau_{rad}$ its mean-lifetime (using 1.05~Myr for ${}^{26}$Al and 3.78~Myr for ${}^{60}$Fe \citep{Rug09}, and $t$ the time. The object formation time $t_F$ is accounted for in the model through the decay of radioactive isotopes that occurs during accretion. No additional accretional heating is accounted for, though. Decay during the formation time $t_F$ results in a decrease of each nuclide initial abundance in the simulations:
\begin{equation}
X_{rad}(t_F)=X_{rad}(0)~e^{-t_F/\tau_{rad}},
\end{equation}
with $X_{rad}(0)$ the initial mass fraction of each nuclide. We consider the initial nuclide ratios ${}^{26}$Al/${}^{27}$Al~=~5~$\times$~10$^{-5}$ \citep{Mac95} and ${}^{60}$Fe/${}^{56}$Fe~=~1~$\times$~10$^{-8}$ \citep{Spi11}, and mass fractions from \citet{Was88} for CV chondrites. 

\subsection{Sensivity to parameters}

Although the model has many free parameters, only a few can actually strongly affect the resulting thermal history, such as the formation time of the object, its composition or the thermal conductivity. We also point out that the object's size plays an important role in its thermal evolution, owing to the surface to volume ratio. Figure \ref{formation} represents the influence of the formation time, through the abundance of radiogenic nuclides. It shows the evolution of the object's central temperature as a function of time after formation, for different formation times. The central temperature corresponds to the maximum temperature achieved inside the object: since the surface is in thermal equilibrium with the surrounding nebula at these stages of evolution, the heating due to radioactive decay is balanced by a {\bf cold load} coming from the surface. Figure \ref{profile} displays the time evolution of the temperature profile within the body, calculated from a formation delay of 3 Myr. This figure shows that, at a given epoch of its evolution and for the adopted set of input parameters, the condition that the temperature profile must be in the 22--47 K range can be satisfied in almost the whole body, thus enabling the release of N$_2$ from this zone. Only the very close subsurface (down to a few hundreds meters deep) remains at lower temperatures, but this region should be rapidly depleted in volatiles once the comet follows subsequent orbits closer to the Sun.

The combined effects of composition and porosity are presented in Figure \ref{composition}. For those simulations, we also considered a plausible formation time of 3~Myr and a fixed density of 700~kg~m$^{-3}$. Figure \ref{conductivity} represents the influence of thermal conductivity for the standard parameters previously mentioned. All together, these simulations show that it is possible to heat planetesimals at temperatures high enough to simultaneously imply the loss of N$_2$ and the preservation of CO, between 22 and 47~K.}

\section{Discussion}

{\bf Our computations show that, under certain sets of conditions and evolutionary paths, initial nitrogen-rich cometesimals similar to Triton and Pluto may have evolved into nitrogen-depleted comets. The proposed scenario of selective trapping in clathrates in the nebula followed by outgassing due to short-lived radiogenic heating thus reconciles present-day observations with solar system formation models.} In our scenario, the composition of comets would have evolved as a result of radiogenic decay of short-lived nuclides. In the case of large objects like Pluto or Triton, the thermal evolution is more complex and involves partial or total physical differentiation. Although this significant processing could involve a substantial loss of volatiles, Pluto and Triton are actually large enough for most volatiles to be retained at the surface and in the atmosphere by gravity, as shown by Schaller \& Brown (2007). Moreover, if accretion of these bodies were occurring in a cold environment and were slow--so that radiative removal of the impact heat were effective-- then Triton, Pluto and other large KBOs could have retained their volatiles and accreted fairly cold (see discussion by Canup \& Ward (2002) in the context of the accretion of Ganymede and Callisto).

Our model predicts that comets should contain Xe/H$_2$O ($\sim$3.6 $\times$ 10$^{-7}$) and Kr/H$_2$O ($\sim$5.0 $\times$ 10$^{-6}$), ratios close to solar Xe/O and Kr/O, respectively. In contrast,  the Ar/H$_2$O ratio is predicted to be equal to $\sim$1.7 $\times$ 10$^{-5}$ in comets, a value about 300 times lower than solar Ar/O. In our model, the Ar depletion (compared to solar) in comets results from the poor propensity of this element to be trapped in different MG clathrates formed in the nebula. Because of this property, Ar was accreted by cometesimals only at low nebular temperature ($\sim$20 K) and in pure condensate form. As a consequence, {\bf the radiogenic heating above 22 K triggers the dissociation of clathrates and the outgassing of N$_2$ and sublimation of Ar.}

Measuring the noble gas abundances in comets might require in situ measurements via a high sensitivity mass spectrometer or ultraviolet spectroscopy. The ALICE ultraviolet spectrometer and the ROSINA instrument should put upper limits on the noble gas abundances in the coma of comet 67P/Churyumov-Gerasimenko (Balsiger et al. 2007; Stern et al. 2007) {\bf that will probably help understand} the formation conditions of comets in the primordial nebula. However, the measurement of solar or subsolar noble gas abundances in comets will require a new generation of instruments such as the cryotrap, which is part of the mass spectrometer MASPEX developed in the context of the PRIME spacecraft project (Young et al. 2010).

\acknowledgements
O. M. and J.-M. P. acknowledge support from CNES, and J. L. from the Distinguished Visiting Scientist Program at JPL. We acknowledge an anonymous Referee for his constructive remarks.

%
%

\clearpage

\begin{table}
\caption[]{Adopted set of parameters for the Kihara potential}
\begin{center}
\begin{tabular}{lcccc}
\hline
\hline
Mol.   		& $\sigma$ (\AA)	& $ \epsilon/k_B$ (K)	& $a$ (\AA) 	& Reference	\\
\hline
CO            		& 3.1515     		& 133.61     			& 0.3976 		& Mohammadi et al. (2005) 		\\
CO$_2$		& 2.9681			& 169.09				& 0.6805  		& Thomas et al. (2009) 			\\
CH$_4$     	& 3.14393     		& 155.593     			& 0.3834 		& Sloan \& Koh (2008)			\\
N$_2$       	& 3.13512     		& 127.426     			& 0.3526 		& Sloan \& Koh (2008)			\\
H$_2$S		& 3.1558			& 205.85				& 0.36 		& Parrish \& Prausnitz (1972)		\\
PH$_3$		& 3.771			& 275.0				& 0 			& Vorotyntsev \& Malyshev (1998)	\\
Ar              		& 2.9434     		& 170.50     			& 0.184 		& Parrish \& Prausnitz (1972)		\\
Kr             		& 2.9739     		& 198.34     			& 0.230 		& Parrish \& Prausnitz (1972)		\\
Xe              	& 3.32968     		& 193.708     			& 0.2357 		& Sloan \& Koh (2008)			\\
\hline
\end{tabular}
\end{center}
$\sigma$ is the Lennard-Jones diameter, $\epsilon$ is the depth of the potential well, and $a$ is the radius of the impenetrable core. \\
\label{kihara}
\end{table}

\clearpage

\clearpage

\clearpage

\clearpage

\clearpage

\begin{figure}
\begin{center}
\resizebox{\hsize}{!}{\includegraphics{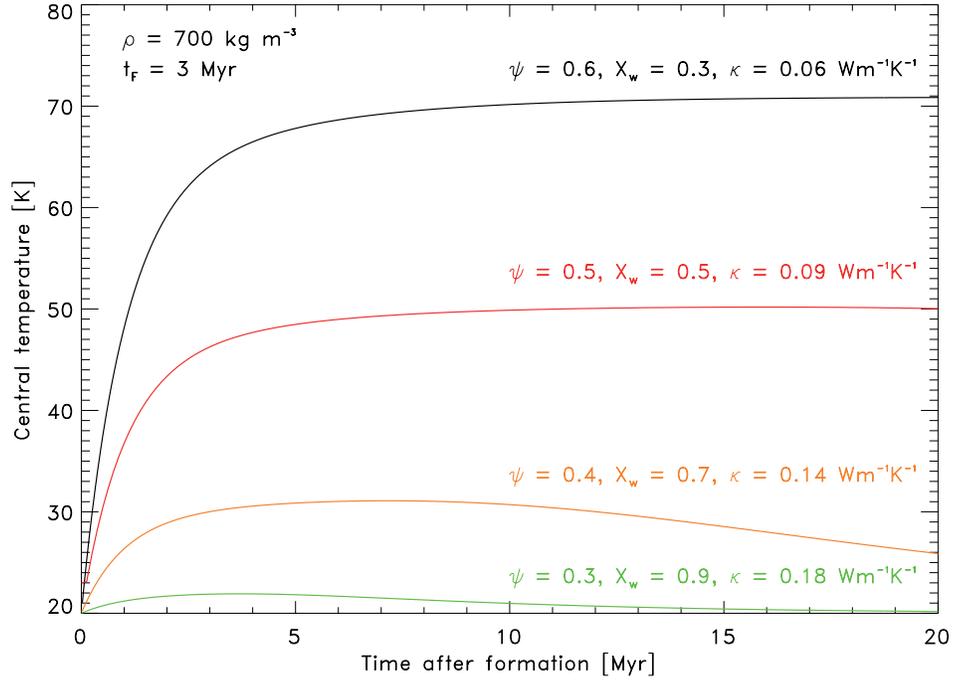}}
\caption{Time evolution of central temperature calculated for different values of the porosity and composition.} 
\label{composition}
\end{center}
\end{figure}

\clearpage

\end{document}